\documentclass[conference]{IEEEtran}
\IEEEoverridecommandlockouts

\usepackage[utf8]{inputenc}

\usepackage{amsmath,amssymb,epsfig,psfrag,cite,subfigure}
\include{macros}
\usepackage{graphicx}
\usepackage{epstopdf}

\usepackage[normalem]{ulem}
\usepackage{pgfplots}
\usepackage{acronym}

\usepackage{bm}

\usepackage{tcolorbox}

\usepackage{accents}
\allowdisplaybreaks
\usepackage{soul}

\usepackage{amsthm}

\usepackage{nicefrac}

\usepackage{lipsum,multicol}

\usepackage{algorithm,balance}
\usepackage{algpseudocode}
\algdef{SE}[SUBALG]{Indent}{EndIndent}{}{\algorithmicend\ }%
\algtext*{Indent}
\algtext*{EndIndent}

\usepackage{enumitem}
\usepackage{mwe}    
\usepackage{epsfig,psfrag}
\usepackage{subfigure}
\usepackage{color}
\usepackage{url}
\usepackage{mathtools,xparse}

\usepackage{gensymb}

\usepackage[multiple]{footmisc}

\usepackage{xr}
\makeatletter
\newcommand*{\addFileDependency}[1]{% argument=file name and extension
  \typeout{(#1)}
  \@addtofilelist{#1}
  \IfFileExists{#1}{}{\typeout{No file #1.}}
}
\makeatother

\newcommand*{\myexternaldocument}[1]{%
    \externaldocument{#1}%
    \addFileDependency{#1.tex}%
    \addFileDependency{#1.aux}%
}

\myexternaldocument{supp}

\usepackage{xcolor}

\newcommand\abs[1]{\big|#1\big|}

\newcommand\norm[1]{\left\lVert #1\right\rVert}

\newcommand\normss[1]{\bigg\lVert #1\bigg\rVert}

\newcommand{\pp}{\boldsymbol{p}}
\newcommand{\vv}{\boldsymbol{v}}
\newcommand{\hh}{\boldsymbol{h}}

\newcommand{\vecc}[1]{ {\rm{vec}}\left(#1\right)  }

\newcommand{\cc}{\boldsymbol{c}}
\newcommand{\dd}{\boldsymbol{d}}

\newcommand{\ggb}{\boldsymbol{g}}

\newcommand{\YY}{\boldsymbol{Y}}
\newcommand{\WW}{\boldsymbol{W}}
\newcommand{\XX}{\boldsymbol{X}}
\newcommand{\ZZ}{\boldsymbol{Z}}

\newcommand{\bb}{\boldsymbol{b}}

\newcommand{\etab}{\boldsymbol{\eta}}

\newcommand{\deltaf}{\Delta_f}

\newcommand{\boldzero}{{ {\boldsymbol{0}} }}

\newcommand{\diag}{ \mathrm{diag}  }
\newcommand{\thn}[1]{ {#1^{\rm{th} } } }

\newcommand{\aab}{\boldsymbol{a}}

\newcommand{\nuhat}{\widehat{\nu}}
\newcommand{\tauhat}{\widehat{\tau}}
\newcommand{\thetabhat}{\widehat{\bm{\theta}}}

\newcommand{\xnm}{x_{n,m}}
\newcommand{\ynm}{y_{n,m}}
\newcommand{\znm}{z_{n,m}}

\newcommand{\Ts}{T_s}
\newcommand{\Tcp}{T_{\rm{cp}}}

\newcommand{\nris}{N_{\rm{RIS}}}
\newcommand{\Pt}{P_t}

\newcommand{\complexset}[2]{ \mathbb{C}^{#1 \times #2}  }

\newcommand{\her}{\mathsf{H}}
\newcommand{\trp}{\mathsf{T}}

\newcommand{\conj}{*}

\newcommand{\betab}{\bm{\beta}}

\newcommand{\thetab}{\bm{\theta}}

\newcommand{\thetaaz}{\theta_{\rm{az}}}
\newcommand{\thetael}{\theta_{\rm{el}}}

\newcommand{\brtext}{{\rm{br}}}

\newcommand{\thetabbr}{\thetab^{\brtext}}
\newcommand{\thetabraz}{\thetaaz^{\brtext}}
\newcommand{\thetabrel}{\thetael^{\brtext}}

\newcommand{\dbr}{d_{\brtext}}

\newcommand{\hdet}{ \underset{\mathcal{H}_0}{\overset{\mathcal{H}_1}{\gtrless}} }
\newcommand{\llrlog}{\mathcal{L}^{\rm{log}}}

\newcommand{\Omegab}{\bm{\Omega}}
\newcommand{\omegab}{\bm{\omega}}

\newcommand{\mtcn}{{\mathcal{CN}}}
\newcommand{\Imatrix}{{ \boldsymbol{\mathrm{I}} }}

\newcommand{\rcs}{\sigma_\text{RCS}}

\newcommand{\1}{\boldsymbol{1}}

\newcommand{\bYY}{\widebar{\boldsymbol{Y}}}

\newcommand{\lr}{\mathcal{L}}

\newcommand{\normf}[1]{\left|\left|#1\right|\right|_F}

\newcommand{\tracesmall}[1]{ {{{\rm{tr}}\left( #1 \right)}}  }

\newcommand{\ind}[2]{ \left[#1\right]_{#2}}

\makeatletter \renewcommand\d[1]{\ensuremath{%
		\;\mathrm{d}#1\@ifnextchar\d{\!}{}}}
\makeatother

\makeatletter
\newcommand*\rel@kern[1]{\kern#1\dimexpr\macc@kerna}
\newcommand*\widebar[1]{%
  \begingroup
  \def\mathaccent##1##2{%
    \rel@kern{0.8}%
    \overline{\rel@kern{-0.8}\macc@nucleus\rel@kern{0.2}}%
    \rel@kern{-0.2}%
  }%
  \macc@depth\@ne
  \let\math@bgroup\@empty \let\math@egroup\macc@set@skewchar
  \mathsurround\z@ \frozen@everymath{\mathgroup\macc@group\relax}%
  \macc@set@skewchar\relax
  \let\mathaccentV\macc@nested@a
  \macc@nested@a\relax111{#1}%
  \endgroup
}
\makeatother

\theoremstyle{remark}

\newtheoremstyle{mytheoremstyle} % name
    {\topsep}                    % Space above
    {\topsep}                    % Space below
    {\upshape}                   % Body font
    {.5em}                           % Indent amount
    {\itshape}                   % Theorem head font
    {.}                          % Punctuation after theorem head
    {.5em}                       % Space after theorem head
    {}  % Theorem head spec (can be left empty, meaning ‘normal’)
    
\theoremstyle{mytheoremstyle}

\newtheoremstyle{iremark}
  {\topsep}   % ABOVESPACE
  {\topsep}   % BELOWSPACE
  {\upshape}  % BODYFONT
  {0.2in}       % INDENT (empty value is the same as 0pt)
  {\itshape}  % HEADFONT
  {.}         % HEADPUNCT
  {5pt plus 1pt minus 1pt} % HEADSPACE
  {\thmname{#1}\thmnumber{ \itshape#2}\thmnote{ (#3)}} % CUSTOM-HEAD-SPEC

\theoremstyle{iremark}

\acrodef{RIS}{reconfigurable intelligent surface}
\acrodef{SNR}{signal-to-noise ratio}
\acrodef{ISAC}{integrated sensing and communications}
\acrodef{SLAM}{simultaneous localization and mapping}
\acrodef{ISLAC}{integrated sensing, localization, and communications}
\acrodef{LoS}{line-of-sight}
\acrodef{NLoS}{non-line-of-sight}
\acrodef{AoA}{angle-of-arrival}
\acrodef{AoD}{angle-of-departure}
\acrodef{UE}{user equipment}
\acrodef{NF}{near-field}
\acrodef{BS}{base station}
\acrodef{MCRB}{misspecified Cram\'{e}r-Rao bound}
\acrodef{CRB}{Cram\'{e}r-Rao bound}
\acrodef{LB}{lower bound}
\acrodef{ML}{maximum-likelihood}
\acrodef{MML}{mismatched maximum-likelihood}
\acrodef{DL}{downlink}
\acrodef{UL}{uplink}
\acrodef{MIMO}{multiple-input multiple-output}
\acrodef{MISO}{multiple-input single-output}
\acrodef{SISO}{single-input single-output}
\acrodef{SIP}{shift invariance property}
\acrodef{FIM}{Fisher information matrix}
\acrodef{RMSE}{root mean-squared error}
\acrodef{AWGN}{additive white Gaussian noise}
\acrodef{ADMM}{alternating direction method of multipliers}
\acrodef{LS}{least-squares}
\acrodef{SOC}{second-order cone}
\acrodef{CFO}{carrier frequency offset}
\acrodef{TX}{transmit}
\acrodef{RX}{receive}
\acrodef{CP}{cyclic prefix}
\acrodef{ISI}{intersymbol interference}
\acrodef{PSD}{power spectral density}
\acrodef{OFDM}{orthogonal frequency-division multiplexing}

\newcommand{\mke}[1]{\textcolor{blue}{{#1}}}

\begin{document}
\bstctlcite{IEEEexample:BSTcontrol}

%%%%%%%%%%%%%%%%%% title page information %%%%%%%%%%%%%%%%%%
\title{RIS-Aided NLoS Monostatic Sensing under Mobility and Angle-Doppler Coupling
\thanks{This study is partially supported by Türk Telekom within the framework of 5G and Beyond Joint Graduate Support Programme coordinated by Information and Communication Technologies Authority. This work is also supported by Hexa-X-II, part of the European Union’s Horizon Europe research and innovation programme under Grant Agreement No 101095759 and the Swedish Research Council (VR grant 2022-03007).}}

%%%%%%%%%%%%%%%%%% authors %%%%%%%%%%%%%%%%%%
\author{\IEEEauthorblockN{Mahmut Kemal Ercan\IEEEauthorrefmark{1}, Musa Furkan Keskin\IEEEauthorrefmark{2}, Sinan Gezici\IEEEauthorrefmark{1}, and Henk Wymeersch\IEEEauthorrefmark{2}}
\IEEEauthorblockA{\IEEEauthorrefmark{1} Department of Electrical and Electronics Engineering, Bilkent University, Turkey \\
\IEEEauthorrefmark{2}  Department of Electrical Engineering, Chalmers University of Technology, Sweden \\
E-mail: ercan@ee.bilkent.edu.tr}
\vspace{-0.6cm}
}

\maketitle

%%%%%%%%%%%%%%%%%% abstract %%%%%%%%%%%%%%%%%%
\begin{abstract}
We investigate the problem of \ac{RIS}-aided monostatic sensing of a mobile target under \ac{LoS} blockage  considering a single-antenna, full-duplex, and dual-functional radar-communications \ac{BS}. %The goal is to detect the target and estimate its delay, Doppler, and angle parameters using the echoes of downlink transmissions through the \ac{RIS}-induced \ac{NLoS} path. 
For the purpose of target detection and delay/Doppler/angle estimation, we derive a detector based on the generalized likelihood ratio test (GLRT), which entails a high-dimensional parameter search and leads to angle-Doppler coupling. % (i.e., slow-time phase shifts due to mobility and RIS beam sweeping cannot be disentangled from each other). 
To tackle these challenges, we propose a two-step algorithm for solving the GLRT detector/estimator in a low-complexity manner, accompanied by a RIS phase profile design tailored to circumvent the angle-Doppler coupling effect. Simulation results verify the effectiveness of the proposed algorithm, demonstrating its convergence to theoretical bounds and its superiority
%and show significant improvements 
over state-of-the-art mobility-agnostic benchmarks.
\end{abstract}

\acresetall 
%%%%%%%%%%%%%%%%%%%%%%%%%%  body  %%%%%%%%%%%%%%%%%%%%%%%%%%
\vspace{-5mm}
\section{Introduction}\label{sec_intro}

%\subsection{Literature Review}

%Reconfigurable intelligent surface (RIS) is an emerging technology that can be used to improve performance of communication and sensing systems \cite{ISAC_RIS_WCM_2023,ISAC_RIS_2023_SPM}. 
\Acp{RIS} have emerged as a transformative technology towards 6G with the promise of significantly boosting communication rates and expanding coverage through the ability to overcome signal blockages, especially in mmWave and sub-THz systems \cite{RIS_tutorial_2021,RIS_THz_2021,RIS_6g_commag_2021}. 
\Acp{RIS} also have benefits to improve sensing capabilities \cite{ISAC_RIS_WCM_2023,ISAC_RIS_2023_SPM}, thereby supporting \ac{ISAC} \cite{Fan_ISAC_6G_JSAC_2022,chepuri2023integrated,Sankar_Joint_2021}. 
%
%Another disruptive technology for 6G wireless networks is \ac{ISAC}, which aims to deliver both functionalities simultaneously by leveraging the existing wireless infrastructure, hardware and spectral resources \cite{Fan_ISAC_6G_JSAC_2022}. Thanks to these advancements, \ac{ISAC} systems employing \acp{RIS} have become popular in recent years 
%
%\cite{ISAC_RIS_WCM_2023,ISAC_RIS_2023_SPM}, with 
Applications include monostatic \cite{Zhang_MetaRadar_2022}, bistatic \cite{foundations_RIS_radar_TSP_2022} and multistatic \cite{wei_target_2021} sensing, as well as \ac{SLAM} \cite{hyowon_slam_ris_2023}. \ac{RIS}-aided sensing brings several important advantages over conventional sensing without RIS, such as \ac{RIS} phase profile design for illumination of desired sectors \cite{isac_beamform_TWC_2023,liu_joint_2022}, \ac{NLoS} sensing in the presence of obstacles \cite{NLOS_RIS_sensing_CRB_TSP_2023,esmaeilbeig2022cramer,aubry_reconfigurable_2021}, \ac{SNR} enhancement \cite{RIS_mono_ICASSP_2022} and high-accuracy angle estimation via large aperture of \ac{RIS} \cite{RIS_SPM_2022,grossi_radar_2023}.

\begin{figure}[t]
	    \centering
	    \includegraphics[width=0.5\columnwidth]{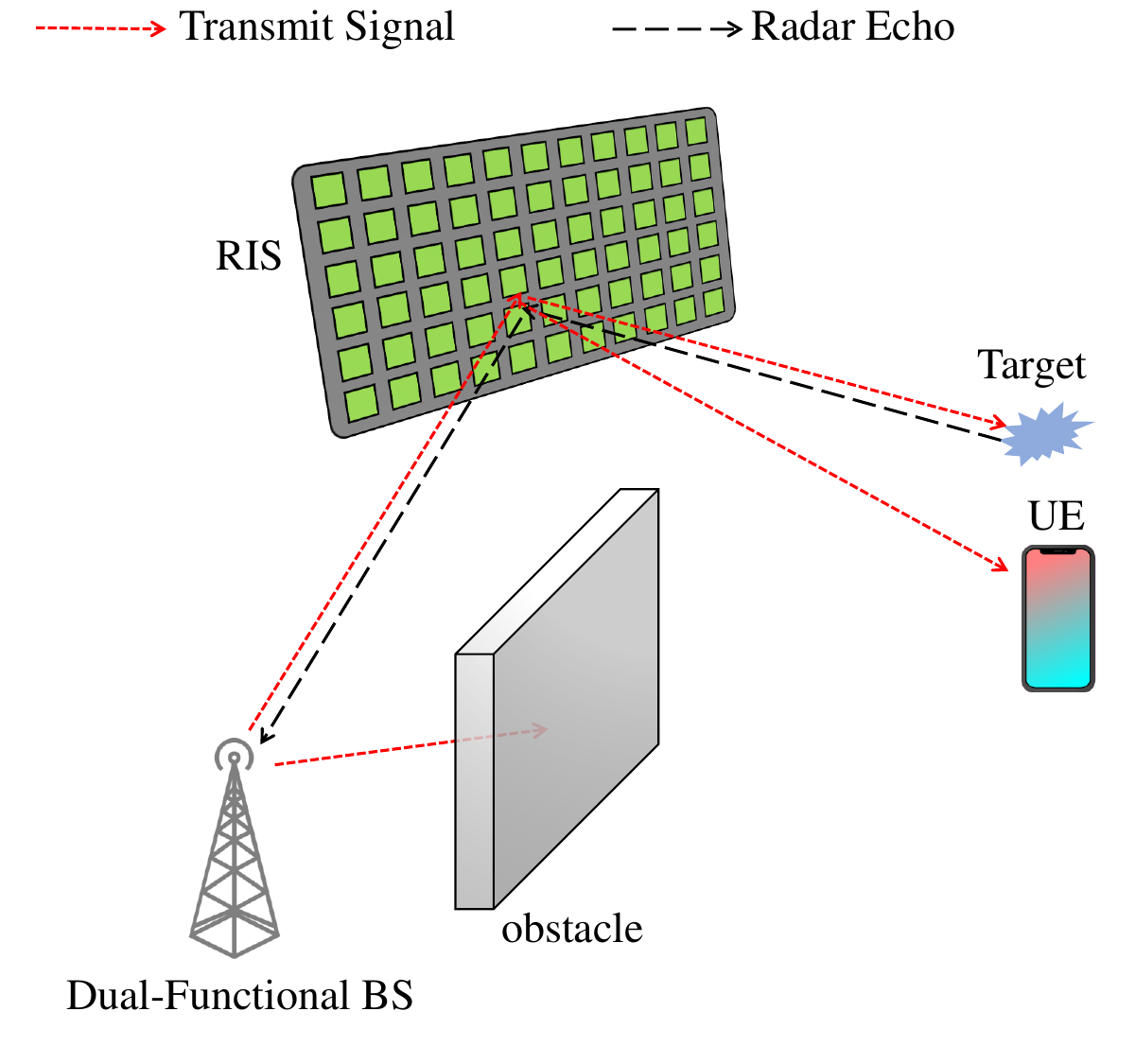}
        \vspace{-5mm}
	    \caption{RIS-aided monostatic sensing scenario under LoS blockage, where a BS communicates with a \acf{UE} in \acf{DL} while detecting a mobile target using the backscattered signals over the RIS path.}
	\label{fig:setup} \vspace{-5mm}
\end{figure}

In the context of RIS-aided sensing, several studies have focused on monostatic configurations \cite{foundations_RIS_radar_TSP_2022,grossi_radar_2023,NLOS_RIS_sensing_CRB_TSP_2023,RIS_sensing_JSAC_2022,aubry_reconfigurable_2021,Esmaeilbeig_IRS_2022,Buzzi_Radar_2021}, particularly due to their attractive properties, including ease of synchronization and utilization of the entire signal (i.e., not only pilots). When the \ac{LoS} path exists between the monostatic radar transceiver and the target, \ac{RIS} can serve to boost the \ac{SNR} via an extra, controlled path towards the target by judicious design of RIS element phases \cite{RIS_mono_ICASSP_2022,Buzzi_Radar_2021,grossi_radar_2023}. Under \ac{LoS} blockage conditions (as shown in Fig.~\ref{fig:setup}), \ac{RIS} can act as an additional anchor to provide high-resolution \ac{AoD}/\ac{AoA} measurements of potential targets \cite{NLOS_RIS_sensing_CRB_TSP_2023,xing2023joint}. Given sufficient bandwidth (e.g., via \ac{OFDM} transmissions \cite{wei_multi_2022,esmaeilbeig2023moving}), the monostatic radar can combine angular information with round-trip delay measurements over the \ac{RIS}-induced \ac{NLoS} path to perform high-accuracy localization of targets in a surveillance area blocked by obstacles \cite{aubry_reconfigurable_2021,xing2023joint}. 

A particular challenge arising in \ac{RIS}-aided \ac{NLoS} monostatic sensing pertains to detection and parameter estimation of \textit{mobile targets}. %In the presence of the \ac{LoS} path, a monostatic \ac{MIMO} radar equipped with a digital receive array can estimate target angles via spatial-domain phase shifts in element-space (i.e., over the array) and Doppler shifts over slow-time phase progressions, using the \ac{LoS} backscattered signal \cite{angleDoppler_MIMO_radar}. However, under \ac{LoS} blockage, 
In particular, target angle information can only be acquired through the \ac{RIS}, which induces beamspace measurements at the radar receiver due to its passive nature (i.e., no sampling mechanism at the \ac{RIS}), in turn requiring several non-parallel \ac{RIS} beams \cite{Keykhosravi2020_SisoRIS}. %This requires the use of multiple different RIS beams (phase profiles) for angle estimation of tracked targets in beamspace since non-parallel RIS phase profile vectors need to be employed to make \ac{AoD} identifiable \cite{Keykhosravi2020_SisoRIS}. 
%In addition, searching over a surveillance area for detection of potential targets necessitates RIS beam sweeping. 
In parallel, target Doppler is measured from slow-time phase shifts across transmissions. 
Consequently, the impact of mobility and RIS beam sweeping on slow-time phase shifts cannot be disentangled from each other, leading to the \textit{angle-Doppler coupling} effect \cite{MIMO_angleDoppler_coupling_2021}. The few studies that considered 
%Even though 
RIS-aided sensing
%has been investigated in the literature, only a few studies have focused on sensing 
of mobile targets \cite{aubry_reconfigurable_2021,esmaeilbeig2022cramer,wei_multi_2022,esmaeilbeig2023moving,Esmaeilbeig_IRS_2022} focused mainly on RIS phase profile optimization and waveform design, omitting the algorithmic aspects. %In those studies, however, the major effort has been devoted to RIS phase profile optimization and waveform design. 
Hence, the two compelling problems remain unexplored in \ac{RIS}-aided \ac{NLoS} monostatic sensing research: \textit{(i)} detection and four-dimensional (range-Doppler-azimuth-elevation) estimation of mobile targets, and \textit{(ii)} the accompanying problem of angle-Doppler coupling.  

In this paper, we tackle the problem of \ac{RIS}-aided \ac{NLoS} monostatic sensing of a mobile target using the echoes of downlink signals transmitted by a dual-functional \ac{ISAC} \ac{BS}. We derive the corresponding generalized likelihood ratio test (GLRT) detector, which involves computationally complex four-dimensional parameter search over range-Doppler-azimuth-elevation domains due to angle-Doppler coupling inherent to \ac{RIS}-aided angle estimation under mobility. To address this challenge, we propose to employ repetitive \ac{RIS} phase profiles and devise a novel low-complexity two-step estimator that exploits the proposed temporal design to circumvent the angle-Doppler coupling effect, enabling separate estimation of delay-Doppler and azimuth-elevation. Simulation results reveal the effectiveness of the proposed phase profile design and the low-complexity algorithm in mobile target sensing.

\section{System Model and Problem Formulation}\label{sec_system_model}
In this section, we describe the \ac{RIS}-aided \ac{NLoS} sensing scenario, derive the signal model at the \ac{BS}, and formulate the problem of detection/estimation of a mobile target.

%%%%%%%%%%%%%%%%%%%%%%%%%%%%%%%%%%%%%%%%%%%%%%%%%%%%%%
\subsection{Scenario}
We consider the RIS-aided monostatic sensing scenario under \ac{LoS} blockage as shown in Fig.~\ref{fig:setup}, where a full-duplex dual-functional radar-communications \ac{BS} with a known location performs monostatic sensing using the echoes of \ac{DL} transmissions through the \ac{RIS}-induced \ac{NLoS} path \cite{ISAC_RIS_2022,RIS_sensing_magazine_2023}. The aim is to detect a mobile target in the environment and estimate its parameters (delay, Doppler, angle). The \ac{BS} is equipped with a single \ac{TX} and a single \ac{RX} antenna, while the \ac{RIS} has $\nris$ elements. In addition, the \ac{RIS} is assumed to have a known location and orientation.

%%%%%%%%%%%%%%%%%%%%%%%%%%%%%%%%%%%%%%%%%%%%%%%%%%%%%%

\subsection{Signal Model}

For \ac{DL} communications with a \ac{UE}, the \ac{BS} transmits an OFDM signal with $N$ subcarriers and $M$ symbols, with $\xnm$ denoting the complex data/pilot on the $\thn{n}$ subcarrier of the $\thn{m}$ symbol. The subcarrier spacing and the symbol duration are defined as $\deltaf = 1/T$ and $\Ts = T + \Tcp$, respectively, with $T$ denoting the elementary symbol duration and $\Tcp$ the \ac{CP} duration. 
We assume the existence of a single target in the environment. 
%characterized by ${\alpha_k, \tau_k, \nu_k, \theta}$. 
To prevent \ac{ISI}, the \ac{CP} duration is assumed to be larger than the round-trip delay of the target over the RIS path \cite{OFDM_Radar_Phd_2014,MIMO_OFDM_ICI_JSTSP_2021}. In this setup, the backscattered OFDM signal at the monostatic sensing receiver of the \ac{BS} on the $\thn{n}$ subcarrier of the $\thn{m}$ symbol (after removing the \ac{CP} and switching to frequency-domain) can be written as \cite{NLOS_RIS_sensing_CRB_TSP_2023,esmaeilbeig2022cramer}
\begin{align} \nonumber
    \ynm &=  \alpha \, \underbrace{\aab^\trp(\thetabbr) \Omegab_m \aab(\thetab)}_{{\rm{Target}\textrm{-}\rm{RIS}\textrm{-}\rm{BS}\,\rm{path}}} \,\underbrace{\aab^\trp(\thetab) \Omegab_m \aab(\thetabbr)}_{{\rm{BS}\textrm{-}\rm{RIS}\textrm{-}\rm{Target}\,\rm{path}}} \\ \label{eq_ynm}
    &~~~~\times [\cc(\tau)]_n [\dd(\nu)]_m  \xnm + \znm \,, 
\end{align}
% \begin{align} \nonumber
%     \ynm &= \underbrace{ \alpha \, (\aab^\trp(\thetab) \Omegab_m \aab(\thetabbr))^2 [\cc(\tau)]_n [\dd(\nu)]_m  \xnm}_{\rm{BS-RIS-Target-RIS-BS~paths}} \\
%     \label{eq_ynm}
%     &~~+ \underbrace{\alphabr \, \aab^\trp(\thetabbr) \Omegab_m \aab(\thetabbr) [\cc(\taubr)]_n   \xnm}_{\rm{BS-RIS-BS~path}}
%     + \znm \,,
% \end{align}
where
\begin{itemize}
    \item $\thetabbr = [\thetabraz, \thetabrel]^\trp$ is the known \ac{AoA} from the \ac{BS} to the \ac{RIS}. 
    %The azimuth angle is the counter-clockwise angle in x-y plane starting from x-axis, and the elevation angle is the angle between the x-y plane and z-axis component.
    
    \item $\thetab = [\thetaaz, \thetael]^\trp$ is the \ac{AoD} from the \ac{RIS} to the target.
    %defined similar to $\thetabbr$.
    
    \item $\alpha$ is the complex channel gain including the effects of \textit{(i)} two-way attenuation over the forward BS-RIS-target path and the backward target-RIS-BS path, and \textit{(ii)} radar-cross-section (RCS) of the target, which can be calculated as \cite{aubry_reconfigurable_2021}
\begin{equation} \label{eq:alphak}
    \abs{\alpha} = \sqrt{\frac{\Pt G_b^2 G^2  F^2(\thetab) F^2(\thetabbr) d_x^2 d_y^2 \lambda^2 \rcs}{(4\pi)^5 \dbr^4  d^4}} \,.
\end{equation}
    Here, $\Pt$ is the transmit power, $G_b$ is the BS antenna gain, $G$ is the antenna power gain of an RIS patch, $F(\thetab) = (\cos{\thetael})^{0.285}$ is the normalized RIS power radiation pattern \cite[Eq.~(3)]{LOS_NLOS_NearField_2021}, $d_x$ and $d_y$ represent the RIS element spacing along the local horizontal and vertical axes, respectively, $\lambda$ denotes the carrier wavelength, $\rcs$ is the target RCS, and, $\dbr$ and $d$ denote the BS-RIS and RIS-target distances, respectively.

    \item $\tau = 2(\dbr + d)/c$ and $\nu$ denote the round-trip delay and Doppler shift of the BS-RIS-target path, respectively, with $c$ denoting the speed of propagation.
    
    \item $\aab(\thetab) \in \complexset{\nris}{1}$ denotes the array steering vector at the \ac{RIS}, given as follows:
    \begin{equation}
        \ind{\aab(\thetab)}{i} = e^{j\frac{2\pi}{\lambda}\cos{\thetael}(\cos{\thetaaz}p_{x,i }+\sin{\thetaaz}p_{y,i} )} ,
    \end{equation}
    where $p_{x,i }$ and $p_{y,i }$ denote the position of the $\thn{i}$ element in the horizontal and vertical axes of the \ac{RIS} with respect to its center, respectively.

     \item $\Omegab_m = \diag(\omegab_m) \in \complexset{\nris}{\nris}$ is the RIS phase profile for the $\thn{m}$ symbol, containing RIS phase shifts $\omegab_m \in \complexset{\nris}{1}$ on its diagonal.
    
    \item $\cc(\tau) \in \complexset{N}{1}$ is the frequency-domain steering vector as a function of delay $\tau$, given by
    \begin{align}
    \cc(\tau) &=   \left[ 1, \ e^{-j 2 \pi \deltaf \tau}, \ \cdots, \ e^{-j 2 \pi (N-1) \deltaf  \tau} \right]^\trp ~.
    \end{align}
    
    \item $\dd(\nu) \in \complexset{M}{1}$ is the time-domain steering vector as a function of Doppler $\nu$, expressed as
    \begin{align}
    \dd(\nu) &=   \left[ 1, \ e^{j 2 \pi \Ts \nu}, \ \cdots, \ e^{j 2 \pi (M-1) \Ts  \nu} \right]^\trp ~.
    \end{align}

    \item $\znm \sim \mtcn(0, N_0 N \deltaf)$ is the additive complex Gaussian noise, with $N_0$ denoting the noise \ac{PSD}.
\end{itemize}

\begin{comment}
    \begin{figure}[t]
	    \centering
	\includegraphics[width=0.7\columnwidth]{Figures/ris_coordinates.jpg}
	    \caption{Definitions of azimuth and elevation angles.}
	\label{fig:azel_coordinates}
\end{figure}
\end{comment}

To provide a more compact representation of \eqref{eq_ynm}, we define
\begin{align}
    \bb(\thetab) \triangleq \aab(\thetab) \odot \aab(\thetabbr) \in \complexset{\nris}{1} ~,
\end{align}
where $\odot$ denotes the Hadamard (element-wise) product. Accordingly, we have
\begin{align} \label{eq_ynm2}
    \ynm &= \alpha \, (\bb^\trp(\thetab) \omegab_m)^2 [\cc(\tau)]_n [\dd(\nu)]_m  \xnm
 + \znm \,.
\end{align}
We further introduce the collection of RIS phase profiles
\begin{align} \label{eq_gammabb}
    \WW = [ \omegab_0 \, \cdots \, \omegab_{M-1} ] \in \complexset{\nris}{M} \,.
\end{align}
Using \eqref{eq_gammabb} and stacking over $N$ subcarriers and $M$ symbols, the model in \eqref{eq_ynm2} can be expressed in compact matrix form as
\begin{align} \label{eq_yy}
    \YY &= \XX \odot \alpha  \cc(\tau) \Big( \dd^\trp(\nu) \odot \bb^\trp(\thetab) \WW \odot \bb^\trp(\thetab) \WW   \Big)   + \ZZ \,,
\end{align}
where $\YY \in \complexset{N}{M}$ with $[\YY]_{n,m} = \ynm$, $\XX \in \complexset{N}{M}$ with $[\XX]_{n,m} = \xnm$ and $\ZZ \in \complexset{N}{M}$ is the additive noise matrix with $\vecc{\ZZ} \sim \mtcn(\boldzero, N_0 N \deltaf \Imatrix)$. For ease of exposition, we assume that all the symbols are $1$, i.e., $\XX$ is an all-one matrix. In this case, \eqref{eq_yy} becomes
\begin{align} \label{eq_yy2}
    \YY &= \alpha  \cc(\tau) \Big( \dd^\trp(\nu) \odot \bb^\trp(\thetab) \WW \odot \bb^\trp(\thetab) \WW   \Big)   + \ZZ \,.
\end{align}

%%%%%%%%%%%%%%%%%%%%%%%%%%%%%%%%%%%%%%%%%%%%%%%%%%%%%%

\subsection{Problem Formulation}\label{sec_form}
Given the frequency/slow-time observation matrix $\YY$ in \eqref{eq_yy2}, our goal is to detect the presence of the target and estimate its delay $\tau$, Doppler $\nu$ and azimuth-elevation angles $\thetab$. Due to the angle-Doppler coupling over the slow-time domain (manifested through the terms $\dd^\trp(\nu)$ and $\bb^\trp(\thetab) \WW$), this poses a computationally complex problem. To tackle this challenge, we derive a generic GLRT detector and propose a tailor-made RIS phase profile design, along with a low-complexity detector/estimator, to resolve the angle-Doppler coupling.  
%However, it is a highly complex task due to the coupling between Doppler and AOA. Therefore, we must resolve the coupling and provide a high-performance target detector under mobility. %For that purpose, we have developed a low-complexity detector/estimator by exploiting the usage of the RIS phase profile. We have provided a low-complexity detector and estimator in Sec. \ref{sec_det_est}. In detail, we first introduce the repetitive RIS phase profile in Sec. \ref{sec_rep_ris}. Afterward, we calculate the expression for GLRT in Sec. \ref{sec_glrt}. Then, we provide our low-complexity ML estimator in Sec. \ref{sec:estimator}.

%%%%%%%%%%%%%%%%%%%%%%%%%%%%%%%%%%%%%%%%%%%%%%%%%%%%%%
\section{Low-Complexity Detector/Estimator Design under Angle-Doppler Coupling}\label{sec_det_est}

In this section, we first derive the {GLRT} detector for the sensing problem stated in Sec.~\ref{sec_form}, which leads to a high-complexity optimization. Then, we propose a low-complexity two-step estimator that avoids high-dimensional parameter search in solving the GLRT, by exploiting the structure of a tailor-made RIS phase profile matrix $\WW$ in \eqref{eq_yy2}.

%%%%%%%%%%%%%%%%%%%%%%%%%%%%%%

\subsection{GLRT Detector}\label{sec_glrt}

%From that point, we will assume that the BS-RIS-BS parts are eliminated. Hence, we can redefine the received signal:
To derive the GLRT detector, we re-cast \eqref{eq_yy2} as
\begin{align} 
        \YY &= \bYY(\etab) + \ZZ \, \in\complexset{N}{M} ,\label{eq:YY_one_tgt}    
\end{align}
where
\begin{align}\label{eq_byy}
     \bYY(\etab) \triangleq \alpha  \cc(\tau) \Big( \dd^\trp(\nu) \odot \bb^\trp(\thetab) \WW \odot \bb^\trp(\thetab) \WW   \Big) \,,
\end{align}
%\begin{equation} \label{eq:YY_one_tgt}
%    \begin{split}
%        \YY &= \XX \odot \Bigg(\alpha  \cc(\tau) \Big( \dd^\trp(\nu) \odot \bb^\trp(\thetab) \WW \odot \bb^\trp(\thetab) \WW   \Big) \Bigg) + \ZZ \,,\\
 %       &= \bYY(\etab) + \ZZ \, \in\complexset{N}{M} ,
  %  \end{split}
%\end{equation}
with $\etab = [\alpha,\tau,\nu,\thetab]^T$ being the unknown parameter vector. Note that the RIS phase configuration matrix $\WW$ is under our control; hence, we have degrees of freedom to design $\WW$ according to our needs for radar detection and estimation. %In that sense, we have designed repetitive RIS phase profiles to overcome the Doppler-angle coupling. 
The hypothesis testing problem for the model in \eqref{eq:YY_one_tgt} can be formulated as
\begin{align}\label{eq_hypotest}
    \YY = \begin{cases}
	\ZZ,&~~ {\rm{under~\mathcal{H}_0}}  \\
	\bYY(\etab) + \ZZ,&~~ {\rm{under~\mathcal{H}_1}} 
	\end{cases} \,
\end{align}
and the corresponding GLRT can be written as
\begin{equation}\label{eq:likelihood}
    \lr(\YY) = \frac{\max_{\etab}p_1(\YY|\etab)}{p_0(\YY)} \hdet \gamma\, 
\end{equation}
where $\gamma$ is a threshold determined by a preset false alarm probability. 
%Since $\vecc{\ZZ} \sim \mtcn(\boldzero, N_0 N \deltaf \Imatrix)$, the 
%we can write the probability distribution for
In \eqref{eq:likelihood}, the likelihood function for the no-target case $\mathcal{H}_0$ is given by 
\begin{equation} \label{eq:p0}
    p_0(\YY) = \frac{1}{(\pi \sigma^2)^{MN}}\exp{\bigg(-\frac{\normf{\YY}^2}{2\sigma^2} \bigg)}\,,
\end{equation}
where $\normf{\cdot}$ denotes the Frobenius norm and $\sigma^2 = N_0N\deltaf$, while the likelihood function for the target-present case $\mathcal{H}_1$ is expressed as %. Additionally, since we add a constant number for each element of the noise matrix, only the mean changes for the target-present case:
\begin{equation} \label{eq:p1}
    p_1(\YY|\etab) = \frac{1}{(\pi \sigma^2)^{MN}}\exp{\bigg(-\frac{\normf{\YY-\bYY(\etab)}^2}{2\sigma^2} \bigg)}\, .
\end{equation}
%By using these distributions, we can calculate the GLRT,
%\begin{equation}\label{eq:likelihood}
 %   \lr(\YY) = \frac{\max_{\etab}p_1(\YY|\etab)}{p_0(\YY)}\, .
%\end{equation}

Substituting \eqref{eq:p0} and \eqref{eq:p1} into \eqref{eq:likelihood} and simplifying, we obtain the log-likelihood ratio up to an additive and multiplicative constant as follows:
\begin{align} 
    %\begin{split}
        \llrlog(\YY) &= \min_{\etab} \normf{\YY - \bYY(\etab)}^2 - \normf{\YY}^2
        %\\
        %&= \min_{\etab} \{\normf{\bYY(\etab)}^2 - 2\realp{\trace{\YY^H\bYY(\etab)}}\}\,
        .\label{eq:likelihood_log} 
    %\end{split}
\end{align}
In \eqref{eq:likelihood_log}, the optimal channel gain can be estimated in closed-form as a function of delay, Doppler and angle parameters \cite{keskin2021limited}:
\begin{equation} \label{eq:alpha_hat}
    \widehat{\alpha} = \frac{\tracesmall{\left(\cc(\tau)\hh^T(\nu,\thetab)\right)^H\YY}}{\normf{\cc(\tau)\hh^T(\nu,\thetab)}^2}\,,
\end{equation}
where  $\hh(\nu,\thetab) \triangleq \dd(\nu)\odot \WW^T\bb(\thetab) \odot \WW^T\bb(\thetab)$. Plugging \eqref{eq:alpha_hat} %and \eqref{eq:YY_one_tgt} substituted 
into \eqref{eq:likelihood_log} yields
\begin{align} 
    &\llrlog(\YY) = - \normf{\YY}^2 \\ &+\min_{\tau,\nu,\thetab} \normss{\YY -  \frac{\tracesmall{\left(\cc(\tau)\hh^T(\nu,\thetab)\right)^H\YY}}{\normf{\cc(\tau)\hh^T(\nu,\thetab)}} \notag \cc(\tau) \hh^T(\nu,\thetab)}_F^2 
    \,,\label{eq:likelihood_glrt_simplified}
\end{align}
which can be simplified into (up to additive constants)
\begin{equation} \label{eq:glrt_max}
    \llrlog(\YY) = \max_{\tau,\nu,\thetab} \frac{\Big|\cc^\her(\tau)\YY \hh^\conj(\nu,\thetab) \Big|^2}{\norm{\hh(\nu,\thetab)}^2} \,.
\end{equation}
In general, no further simplification is possible; hence, a 4D search is needed to estimate delay, Doppler, and 2D angle.  
%There are several ways to calculate GLRT. One can search over all 4D space to find minimizing parameters $\tau,\nu,\thetab$. Besides in our case, we will first use ML estimator that exploits the repetitive RIS phase profile to estimate delay, Doppler, and angles. Afterward, use these estimates to calculate GLRT given in \eqref{eq:likelihood_glrt_simplified}.

%%%%%%%%%%%%%%%%%%%%

\subsection{Low-Complexity Estimator} \label{sec:estimator}

To avoid the 4D search, we harness the possibility of designing the RIS phase profile $\WW$ in \eqref{eq_byy}. With the proposed design, the observation model can be expressed in a form that allows disentangling angle-Doppler coupling. Then, the  delay-Doppler estimation is described, followed by 2D angle processing and a refinement stage. 
%Finally, a refinement of the estimates and target detection are presented. 

\subsubsection{Repetitive RIS Phase Profile Design} \label{sec_rep_ris}

The RIS phase profile is adjusted at each symbol duration:
\begin{align}
    \WW &= [ \omegab_0 \, \cdots \, \omegab_{M-1} ] \in \complexset{\nris}{M} ,\\
    \bb^\trp(\thetab) \WW &= [ \bb^\trp(\thetab)\omegab_0 \, \cdots \, \bb^\trp(\thetab)\omegab_{M-1} ] \in \complexset{1}{M} . \label{eq:repeat_result}
\end{align}
From \eqref{eq_byy}, we can see that the time-domain part of the signal depends on $\dd(\nu)$ and $\bb^\trp(\thetab) \WW$. We will try to eliminate the effects of coupling between $\thetab$ and $\nu$. To this end, we can repeat RIS profiles consecutively. In that case, we use $L$ profiles and repeat each phase profile $M/L$ times: 
\begin{equation}    \underbrace{\omegab_1\omegab_1\hdots \omegab_1}_{M/L\text{~times}} \, \underbrace{\omegab_2\omegab_2\hdots \omegab_2}_{M/L \text{~times}} \, \hdots \, \underbrace{\omegab_{L} \hdots\omegab_{L}}_{M/L\text{~times}} \,.
\end{equation}
With this design, during each time interval (i.e., $M/L$  times), the phase shift caused by the angles is constant since the RIS profile is constant. Hence, we can eliminate the coupling effect during each time interval. 

%Before mentioning how we can eliminate the coupling for parameter estimation, we will first derive the GLRT for target detection.

%We must estimate the target parameters $\etab$ and calculate GLRT accordingly to detect the target. Hence, in this strategy, we first estimate $\alpha$ by using the ML estimation given in \eqref{eq:alpha_hat}. Secondly, we estimate delay and Doppler by exploiting the repeated RIS phase profile. Afterward, by using the estimations $\tauhat,\nuhat$, we try to estimate the target angles $\thetabhat$. Lastly, we perform 4D iterative search by setting the estimates $\tauhat,\nuhat,\thetabhat$ as the initial point.

\subsubsection{Observation Model with Repetitive RIS Phase Profiles}

We will again use the signal model given in \eqref{eq_byy}. But, to exploit the repetitive RIS phase profile on parameter estimation, we will use another form of \eqref{eq_byy} For that reason, we define
%For convenience, we define
\begin{equation} 
    \ggb^\trp(\thetab)  =\bb^\trp(\thetab) \WW \odot \bb^\trp(\thetab) \WW \,,
\end{equation}
leading to
\begin{equation}
    \YY = \alpha  \cc(\tau) ( \dd^\trp(\nu) \odot \ggb^\trp(\thetab))+ \ZZ . 
\end{equation}
%We will use this definition to apply the estimator.
Using the RIS phase profile with repetition, we can rewrite $\ggb^\trp(\thetab)$ as
 \begin{align}
 %   \YY &= \alpha \cc(\tau) \left( \dd^\trp(\nu) \odot \ggb^\trp(\thetab) \right)\,,\\
    \ggb^\trp(\thetab) &= \underbrace{(\bb^\trp(\thetab)[\omegab_1\hdots\omegab_L])^2}_{\triangleq \ggb_L^\trp(\thetab)\in\complexset{1}{L}} \otimes \1^\trp_{M/L} \,.
\end{align}
Therefore,
\begin{equation} \label{eq:rec_g_d}
    \YY = \alpha \cc(\tau) [ \dd^\trp(\nu) \odot \big(\ggb_L^\trp(\thetab)\otimes \1^\trp_{M/L}\big)]+\ZZ\,.
\end{equation}
Next, we define
\begin{align*}
    \dd_{M/L}(\nu)&\triangleq [ 1\, e^{j2\pi\nu \Ts} \, \hdots \, e^{j2\pi\nu(M/L-1)\Ts} ]^\trp\in\complexset{M/L}{1}\\
    \dd_L(\nu)&\triangleq [ 1 \, e^{j2\pi\nu\frac{M}{L} \Ts} \, \hdots \, e^{j2\pi\nu\frac{M}{L}(L-1)\Ts} ]^\trp\in\complexset{L}{1},
\end{align*}
which allows us to write $\dd(\nu) = \dd_L(\nu) \otimes \dd_{M/L}(\nu)$. Plugging this into \eqref{eq:rec_g_d} yields 
\begin{align} 
   % \begin{split}
    \YY &= \alpha \cc(\tau) \left[\big(\ggb_L(\thetab)\otimes \1_{M/L}\big) \odot \big(\dd_L(\nu) \otimes \dd_{M/L}(\nu)\big) \right]^\trp+\ZZ\,,\notag \\
    &= \alpha \cc(\tau) \left[\big(\ggb_L(\thetab)\odot \dd_L(\nu)\big) \otimes \big(\1_{M/L} \odot \dd_{M/L}(\nu)\big) \right]^\trp+\ZZ\,,\notag \\
    &= \alpha \cc(\tau) \left[\big(\ggb_L(\thetab)\odot \dd_L(\nu)\big) \otimes \big(\dd_{M/L}(\nu)\big) \right]^\trp+\ZZ\,.\label{eq:obser_Y}
   % \end{split}
\end{align}
With this, we are now ready to introduce  $\betab\triangleq\alpha\ggb_L(\thetab)\odot\dd_L(\nu)\in\complexset{L}{1}$, which will be a key to break down the 4D problem. With $\betab$, we have that
\begin{equation}
    \YY = \alpha \cc(\tau) \big( \betab\otimes\dd_{M/L}(\nu) \big)^\trp+\ZZ.
\end{equation}
By breaking up $\YY = [
        \YY_1 \, \cdots \, \YY_L \,
    ]$, 
%We can see that the received signal depends on delay and Doppler for each element of $\betab$. Therefore, we have
%\begin{equation}
 %   \YY = \begin{bmatrix}
  %      \YY_1 & \hdots & \YY_L \,,
   % \end{bmatrix}
%\end{equation}
we have that $\YY_l = \beta_l\cc(\tau)\dd_{M/L}^\trp(\nu)+\ZZ_l\in\complexset{N}{M/L},\, l=1,\hdots, L$.

\subsubsection{Delay-Doppler Estimation}

We estimate $\tau$ and $\nu$ from $\{\YY_l\}_{l=1}^{L}$ by treating $\betab = \begin{bmatrix} \beta_1,  \ldots,\beta_L \end{bmatrix}^\trp$ as  nuisance parameters.
%Observations:
%\begin{equation*}
 %   \YY_l = \beta_l\cc(\tau)\dd_{M/L}^\trp(\nu) + \ZZ_l \in\complexset{N}{M/L}
%\end{equation*}
%for $l=1,\hdots,L$. $\ZZ_l$ has i.i.d. complex gaussian entries: $\ind{\ZZ_l}{n,m} \sim \mtcn(0, N_0 N \deltaf)$
%Estimate $[\betab^\trp \tau \nu]^\trp$ from $\{\YY_l\}_{l=1}^L$
Due to the independence of noise across $l$, %and they have the same noise variance, we can define 
the \ac{ML} estimator is expressed as
\begin{equation}
    \min_{\tau,\nu ,\betab} \sum_{l=1}^L\normf{\YY_l-\beta_l\cc(\tau)\dd_{M/L}^\trp(\nu)}^2 \,.
\end{equation}
Here, $\betab$ can be obtained in closed form as a function of $\tau$ and $\nu$, leading to 
\begin{equation} \label{eq:delayDoppEst}
    (\tauhat,\nuhat) = \arg \max_{\tau,\nu} \sum_{l=1}^L \left| \cc^\her(\tau)\YY_l\dd_{M/L}^\conj(\nu)\right|^2 \,.
\end{equation}
For a fast implementation of \eqref{eq:delayDoppEst}, we can first obtain a coarse estimate and then refine it via gradient ascent. To obtain the coarse estimate, we can evaluate the objective function in \eqref{eq:delayDoppEst} by performing IFFT over the frequency domain and FFT over the time domain for each $\YY_l$, and integrating non-coherently across $l$, which can be written as $\YY_{\tau,\nu}=\sum_{l=1}^L|\text{FFT(IFFT($\YY_l$,2),1)}|^2$. 
Afterwards, we obtain the initial estimates for $(\tau,\nu)$ as the argument that maximizes $\YY_{\tau,\nu}$.  
%or estimate $\tauhat$ and $\nuhat$ sequentially as follows. 
%.{For further speed-ups, %we perform such FFTs, but %instead of picking the 2D point corresponding in delay-%Doppler map, we just 
%Afterwards, we first combine the Doppler domain samples %as $\sum_{i=1}^M [\YY_{\tau,\nu}]_{:,i}$ and pick the %maximizer point as the delay estimate. Then, we 
%estimate the Doppler velocity by selecting the 
%maximizer in the column of $\YY_{\tau,\nu}$ 
%corresponding to the delay estimate. 
Note that since the range-Doppler map is quantized, one should perform fine tuning for better approximation. Therefore, in our application, we first apply the aforementioned fast implementation %with the FFTs 
and then perform gradient ascent over the objective function in \eqref{eq:delayDoppEst} by employing the initial estimates. 
%obtained from the fast implementation as the initial value.
%\begin{itemize}
%    \item For each subinterval ($\YY_l$), perform 2D FFT
%    \item Integrate non-coherently across subintervals.
 %   \item Take the estimates as the corresponding peak point.
%\end{itemize}

\subsubsection{Angle Estimation}
Given delay-Doppler estimates $(\tauhat,\nuhat)$, we now wish to estimate azimuth and elevation angles $\thetab = [\theta_{az}\, \theta_{el}]^\trp$ from the observations (see \eqref{eq:obser_Y}, repeated RIS phase profiles). We find that 
%\begin{equation} \label{eq:obser_Wnoise}
%    \YY = \alpha \cc(\tau) \left[\big(\ggb_L(\thetab)\odot \dd_L(\nu)\big) %\otimes \big(\dd_{M/L}(\nu)\big) \right]^\trp + \ZZ
%\end{equation}
%Let's plug $(\tauhat, \nuhat)$ into \eqref{eq:obser_Wnoise}. 
\begin{equation} \label{eq:obser_Wests}
    \YY = \alpha \cc(\tauhat) \left[\big(\ggb_L(\thetab)\odot \dd_L(\nuhat)\big) \otimes \big(\dd_{M/L}(\nuhat)\big) \right]^\trp + \ZZ.
\end{equation}
Again, $\alpha$ can be solved in closed form, 
%Then, the problem turns into estimating $\alpha$ and $\thetab$ from \eqref{eq:obser_Wests}. Therefore, ML estimator becomes
%\begin{equation}
%\begin{split}
 %   (\hat{\alpha},\hat{\thetab}) &= \arg \min_{\alpha,\thetab} \sum_{l=1}^L\Bigg|\Bigg|\YY_l\\
  %  &-\alpha \cc(\tauhat) \left[\big(\ggb_L(\thetab)\odot \dd_L(\nuhat)\big) \otimes \big(\dd_{M/L}(\nuhat)\big) \right]^\trp\Bigg|\Bigg|_F^2
%\end{split}
%\end{equation}
%Eliminating the $\alpha$ terms and simplifying, the \underline{ML estimator for angle} can be written as follows
so that the ML estimator becomes
\begin{equation} \label{eq:angleEst}
    \thetabhat = \arg \max_{\thetab} \frac{\left|\cc^\her(\tauhat)\YY\left(\left(\ggb_L(\thetab)\odot\dd_L(\nuhat)\right)\otimes\dd_{M/L}(\nuhat)\right)^\conj \right|^2}{\norm{\ggb_L(\thetab)}^2},
\end{equation}
which involves a 2D search over $\thetab$. %\mke{At this search, a grid of points with 2-degree step size in azimuth-elevation is created. More specifically, the selected grid points are $\thetaaz\gets \{-90\degree ,-88\degree,\hdots,90\degree\}$ and $\thetael\gets \{0\degree ,2\degree,\hdots,90\degree\}$. We choose the maximum point and perform gradient search by using the objective of \eqref{eq:angleEst}.}

\subsubsection{Refinement and Target Detection}

We use the estimates $\tauhat,\nuhat,\thetabhat$ to initialize a 4D gradient ascent on % to solve the maximization problem in 
\eqref{eq:glrt_max},
%, \mke{ which is equivalent to }
%\begin{equation} \label{eq:4D_obj}
%    (\tauhat, \nuhat, \thetabhat)=\arg \max_{\tau,\nu,\thetab} \frac{\left|\cc^\her(\tauhat) \YY \Big(\ggb(\thetabhat) \odot \dd(\nuhat)\Big)^\conj\right|^2}{\norm{\ggb(\thetabhat) \odot \dd(\nuhat)}^2} .
%\end{equation}
%\mke{As we will discuss in the Section \ref{sec_results}, it is very likely for estimations to converge at a local maximum. To overcome such case, we perform a grid search around the estimates obtained from \eqref{eq:delayDoppEst} and \eqref{eq:angleEst}. The grid points are only selected in Doppler domain and are selected as $\ggb_\text{dopp} \gets -r_\text{dopp},-r_\text{dopp}+d_\text{dopp} ,\hdots,r_\text{dopp}$ where $r_\text{dopp}=50, d_\text{dopp}=0.5$. We evaluate the objective of \eqref{eq:4D_obj} for each grid point for once and perform gradient search initialized with the maximizer point. The complete algorithm for the joint estimator is given in Algorithm \ref{alg:joint}.}
 after which %After estimating the delay-Doppler-angle parameters as $(\tauhat, \nuhat, \thetabhat)$, we plug them 
 we plug the refined estimates into the original GRLT expression in \eqref{eq:glrt_max}, which then becomes%yielding the following detection test
\begin{equation} \label{eq:GLRT_4D}
    \frac{\left|\cc^\her(\tauhat) \YY \Big(\ggb(\thetabhat) \odot \dd(\nuhat)\Big)^\conj\right|^2}{\norm{\ggb(\thetabhat) \odot \dd(\nuhat)}^2} \hdet \gamma ~.
\end{equation}
\section{Simulation Results} \label{sec_results}

\subsection{Scenario and Parameters}
In the simulation setup, the center of the RIS is located at the origin, pointing towards the z-axis. We consider a point target $10\,$m away from the RIS with angle $\thetab = (45\degree,60\degree)$, corresponding to $\pp_\text{tgt} = \begin{bmatrix}3.5355 & 3.5355 & 8.6603 \end{bmatrix}^\trp$ in Cartesian coordinates. The target has a velocity vector of { $\vv=[30,0,30]$ m/s, which results in a relative speed of $\nu=-36.59$ m/s with respect to the RIS.} The BS is located at $\pp_\text{BS} = \begin{bmatrix}  -3.0618& 3.0618 & 2.5\end{bmatrix}^\trp$,  $5\,$m away from the RIS with angle $\thetabbr = (135\degree,30\degree)$.  %which corresponds to $\pp_\text{BS} = \begin{bmatrix}  -3.0618& 3.0618 & 2.5\end{bmatrix}^\trp$. 
\begin{comment}
    \begin{figure}[t]
    \centering
    \includegraphics[width=0.9\columnwidth]{Figures/scenario_cross.jpg}
            %\vspace{-0.5cm}
    \caption{Studied scenario. Yellow arrow shows the RIS direction. A target 10m from the RIS with angle $\thetab = (45\degree,60\degree)$, and relative speed $\nu=-1.22$ m/s, and the BS is located 5m away from the RIS with $\thetabbr = (135\degree,30\degree)$.}
    \label{fig:scenario}
\end{figure}
\end{comment}
%We set the BS gain to 18.06 dB, considering directional transmission to/from the RIS. %We define the BS antenna gain similar to phased array antennas to choose realistic parameters. Commonly, phased array antennas have isotropic elements. In our scenario, we consider the case where BS has $N_\text{BS}=64$ elements and has an antenna gain of $G_b = 10\log_{10} (N_\text{BS})=18.06\,$dB. %Similarly, we choose the same antenna gain. 
%The cyclic prefix is set to avoid ISI. 
We generate RIS phase profiles to scan the region behind the obstacle (see Fig.~\ref{fig:setup}), which is a quarter sphere. In other words, we create equally spaced angular beams in the region of $\thetaaz\in[-90\degree,90\degree]$ and $\thetael\in[0\degree,90\degree)$. 
In addition, $\lambda=1.07\, \text{cm}$,  $\rcs= 2 \,\text{m}^2$, $\nris=21\times 21$, RIS element spacing $\lambda/4$, $G_b=18.06\, \text{dB}$, $G= 0\, \text{dB}$, $N= 1024$ subcarriers with spacing $\Delta_f=120 \, \text{kHz}$ and  $M=1120$ symbols. The noise power spectral density is $N_0= -174 \,\text{dBm/Hz}$ and the receiver  noise figure is 8 dB.

\begin{comment}

%The other test parameters are given in Table \ref{tab:params}.

\begin{table}%h]
    \centering
    \caption{Simulation Parameters}
    \begin{tabular}{|c|c|}
    \hline
        % Radar parameters
        $P_t$ &  10 to 40 dBm\\
        \hline
        $G_b$ & 18.06 dB\\
        \hline
        $f_c$ & 28 GHz\\
        \hline
        % Target related
        $\rcs$ & 2 $m^2$ \\
        \hline
        % RIS parameters
        $\nris$ & $21^2$\\
        \hline
        $d_\text{RIS}$ & $\lambda/4 = 2.7mm$ \\
        %\hline
        %$F(\thetab)$ & $(\cos{\thetab})^{0.285}$\\
        \hline
        $G$ & 0 dB \\
        \hline
        % OFDM parameters
        $N$ & 1024\\ % number of subcarrier
        \hline
        $M$ & 1120 \\ %number of symbols
        \hline
        $\Delta_f$ & $120 \, \rm{kHz}$ \\ %subcarrier spacing
        \hline
        $T$ & $1/\deltaf=8.33\mu s$\\
        \hline
        % Noise parameters
        $N_0$ & -174 dBm/Hz \\
        \hline
        Noise figure & 8dB \\
        \hline
        
        \hline
    \end{tabular}\label{tab:params}
\end{table}
\end{comment}

We vary the BS transmit power $P_t$ and the number of RIS phase repetitions $M/L$ for the considered scenario and perform the following analyses. We run $1000$ Monte-Carlo tests for a given $P_t$ and $M/L$ to evaluate the performance of the proposed algorithm (termed `Joint') in
%perform 4D ML estimations as discussed in 
Sec.~\ref{sec:estimator}. To study detection performance, we also need the probability of false alarms. Therefore, we perform the same procedure for noise-only signals with the same number of tests. As a benchmark, we consider an estimator that ignores the Doppler (denoted by `DI' for Doppler-ignorant), similar to the studies \cite{Zhang_MetaRadar_2022,Sankar_Joint_2021,foundations_RIS_radar_TSP_2022}. %NLOS_RIS_sensing_CRB_TSP_2023
The DI estimator is similar to the proposed estimator, but simply assumes $\nu=0$.

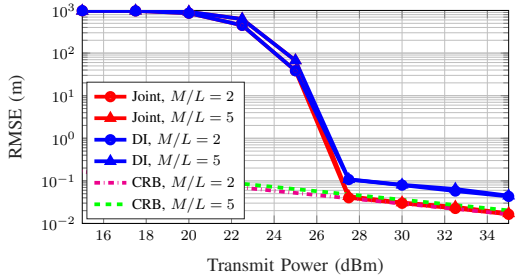
\begin{figure}[t]
    \centering
\begin{tikzpicture}[scale=0.8\columnwidth/10cm,font=\footnotesize]
\begin{axis}[%
width=8 cm,
height=4 cm,
at={(1.011in,0.642in)},
scale only axis,
xmin=15,
xmax=35,
xlabel style={font=\color{white!15!black}},
xlabel={Transmit Power (dBm)},
ymode=log,
ymin=0.01,
ymax=1004.28398200382,
yminorticks=true,
ylabel style={font=\color{white!15!black}},
ylabel={RMSE (m)},
axis background/.style={fill=white},
xmajorgrids,
ymajorgrids,
yminorgrids,
legend style={at={(0.01,0.02)}, anchor=south west, legend cell align=left, align=left, draw=white!15!black},
legend columns=1
%legend pos=south west,
%legend style={legend cell align=left, align=left, draw=white!15!black}
]
\addplot [color=red, line width=2.0pt, mark=o, mark options={solid, red}]
  table[row sep=crcr]{%
15	987.817602156074\\
17.5000000000000	977.304845294985\\
20	863.596033667061\\
22.5000000000000	448.703737964357\\
25	38.5434497180804\\
27.5000000000000	0.0400970392720214\\
30	0.0297503827178094\\
32.5000000000000	0.0225027612963736\\
35	0.0163533215421436\\
};
\addlegendentry{Joint, $M/L=2$}

\addplot [color=red, line width=2.0pt, mark=triangle, mark options={solid, red}]
  table[row sep=crcr]{%
15	978.425693520190\\
17.5000000000000	984.936052298752\\
20	928.276982524711\\
22.5000000000000	629.672767009790\\
25	66.9839390538552\\
27.5000000000000	0.0434965308599372\\
30	0.0308968617139329\\
32.5000000000000	0.0245196292331555\\
35	0.0172217247091351\\
};
\addlegendentry{Joint, $M/L=5$}

\addplot [color=blue, line width=2.0pt, mark=o, mark options={solid, blue}]
  table[row sep=crcr]{%
15	988.793452677497\\
17.5000000000000	978.172001595890\\
20	864.720889293227\\
22.5000000000000	449.132243856292\\
25	38.5722675782381\\
27.5000000000000	0.107489232163813\\
30	0.0794511195804011\\
32.5000000000000	0.0576550634356775\\
35	0.0430885406938715\\
};
\addlegendentry{DI, $M/L=2$}

\addplot [color=blue, line width=2.0pt, mark=triangle, mark options={solid, blue}]
  table[row sep=crcr]{%
15	979.613023778122\\
17.5000000000000	986.342924243313\\
20	929.805933432445\\
22.5000000000000	630.079828776774\\
25	67.0125412716790\\
27.5000000000000	0.108774842716872\\
30	0.0806123741198739\\
32.5000000000000	0.0642897091726969\\
35	0.0448354085415060\\
};
\addlegendentry{DI, $M/L=5$}

\addplot [color=magenta, dashdotted, line width=2.0pt]
  table[row sep=crcr]{%
10	0.292930752461633\\
12.5	0.219667075006378\\
15	0.164727067528281\\
17.5	0.123527874059774\\
20	0.0926328374485735\\
22.5	0.0694648283967199\\
25	0.0520912725669731\\
27.5	0.0390629436547317\\
30	0.0292930752461633\\
32.5	0.0219667075006377\\
35	0.0164727067528281\\
37.5	0.0123527874059774\\
40	0.00926328374485736\\
};
\addlegendentry{CRB, $M/L=2$}

\addplot [color=green, dashed, line width=2.0pt]
  table[row sep=crcr]{%
10	0.360454869946611\\
12.5	0.270303019698646\\
15	0.202698669237092\\
17.5	0.15200255830029\\
20	0.113985838273107\\
22.5	0.0854773200669084\\
25	0.0640989473474315\\
27.5	0.0480674294401448\\
30	0.0360454869946611\\
32.5	0.0270303019698647\\
35	0.0202698669237092\\
37.5	0.015200255830029\\
40	0.0113985838273106\\
};
\addlegendentry{CRB, $M/L=5$}

\end{axis}
\end{tikzpicture}%
            \vspace{-0.3cm}
    \caption{RMSE of range estimation of joint and Doppler-ignorant (DI) estimator for $M/L=2$ and $M/L=5$ with CRB comparison.}
    \label{fig:rmse_range} \vspace{-5mm}
\end{figure}

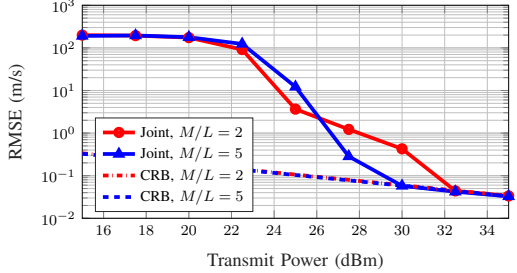
\begin{figure}[t]
    \centering
\begin{tikzpicture}[scale=0.8\columnwidth/10cm,font=\footnotesize]
\begin{axis}[%
width=8 cm,
height=4 cm,
at={(1.011in,0.642in)},
scale only axis,
xmin=15,
xmax=35,
xlabel style={font=\color{white!15!black}},
xlabel={Transmit Power (dBm)},
ymode=log,
ymin=0.01,
ymax=1000,
yminorticks=true,
ylabel style={font=\color{white!15!black}},
ylabel={RMSE (m/s)},
axis background/.style={fill=white},
xmajorgrids,
ymajorgrids,
yminorgrids,
legend pos=south west,
legend style={legend cell align=left, align=left, draw=white!15!black}
]
\addplot [color=red, line width=2.0pt, mark=o, mark options={solid, red}]
  table[row sep=crcr]{%
15	199.243780158084\\
17.5000000000000	194.163167940730\\
20	174.228060997417\\
22.5000000000000	91.3895720013635\\
25	3.65264668570572\\
27.5000000000000	1.22272600933260\\
30	0.427381773468185\\
32.5000000000000	0.0439498826946753\\
35	0.0334928255770541\\
};
\addlegendentry{Joint, $M/L=2$}

\addplot [color=blue, line width=2.0pt, mark=triangle, mark options={solid, blue}]
  table[row sep=crcr]{%
15	190.855917549964\\
17.5000000000000	195.664218994481\\
20	178.050278183231\\
22.5000000000000	123.834231220601\\
25	12.1888612815114\\
27.5000000000000	0.283925327056736\\
30	0.0577413244336539\\
32.5000000000000	0.0421210208191051\\
35	0.0327317648156685\\
};
\addlegendentry{Joint, $M/L=5$}

\addplot [color=red, dashdotted, line width=2.0pt]
  table[row sep=crcr]{%
15	0.336285841175641\\
17.5000000000000	0.252178804979125\\
20	0.189107425571529\\
22.5000000000000	0.141810563378049\\
25	0.106342920299630\\
27.5000000000000	0.0797459401357261\\
30	0.0598010187254517\\
32.5000000000000	0.0448444376542461\\
35	0.0336285841178686\\
};
\addlegendentry{CRB, $M/L=2$}

\addplot [color=blue, dashed, line width=2.0pt]
  table[row sep=crcr]{%
15	0.327917487523110\\
17.5000000000000	0.245903425032826\\
20	0.184401554487220\\
22.5000000000000	0.138281657902165\\
25	0.103696614517354\\
27.5000000000000	0.0777614907540377\\
30	0.0583128916254873\\
32.5000000000000	0.0437284997595780\\
35	0.0327917487524856\\
};
\addlegendentry{CRB, $M/L=5$}

\end{axis}
\end{tikzpicture}%
            \vspace{-0.3cm}
    \caption{RMSE of velocity estimation for $M/L=2$ and $M/L=5$ with CRB comparison.}
    \label{fig:rmse_dopp} \vspace{-5mm}
\end{figure}

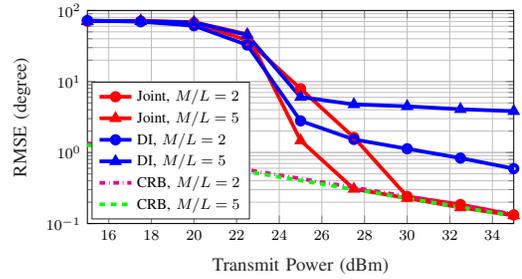
\begin{figure}[t]
    \centering
\begin{tikzpicture}[scale=0.8\columnwidth/10cm,font=\footnotesize]
\begin{axis}[%
width=8 cm,
height=4 cm,
at={(1.011in,0.642in)},
scale only axis,
xmin=15,
xmax=35,
xlabel style={font=\color{white!15!black}},
xlabel={Transmit Power (dBm)},
ymode=log,
ymin=0.1,
ymax=100,
yminorticks=true,
ylabel style={font=\color{white!15!black}},
ylabel={RMSE (degree)},
axis background/.style={fill=white},
xmajorgrids,
ymajorgrids,
yminorgrids,
legend style={at={(0.01,0.02)}, anchor=south west, legend cell align=left, align=left, draw=white!15!black},
]
\addplot [color=red, line width=2.0pt, mark=o, mark options={solid, red}]
  table[row sep=crcr]{%
15	70.5930835016725\\
17.5000000000000	71.0161812614030\\
20	67.1018382466137\\
22.5000000000000	38.0729608196129\\
25	7.91365160778588\\
27.5000000000000	1.63155432939552\\
30	0.240477726544964\\
32.5000000000000	0.185167383429691\\
35	0.132071981438653\\
};
\addlegendentry{Joint, $M/L=2$}

\addplot [color=red, line width=2.0pt, mark=triangle, mark options={solid, red}]
  table[row sep=crcr]{%
15	70.7546296799507\\
17.5000000000000	71.6114050228997\\
20	68.5037025189133\\
22.5000000000000	46.2267850109860\\
25	1.47753230405976\\
27.5000000000000	0.306885990630404\\
30	0.228061041944020\\
32.5000000000000	0.168278414327498\\
35	0.129785275589106\\
};
\addlegendentry{Joint, $M/L=5$}

\addplot [color=blue, line width=2.0pt, mark=o, mark options={solid, blue}]
  table[row sep=crcr]{%
15	72.2988600982693\\
17.5000000000000	69.4376473843881\\
20	61.0812025381852\\
22.5000000000000	32.6703801321527\\
25	2.78856957901046\\
27.5000000000000	1.52831426000795\\
30	1.12957533201903\\
32.5000000000000	0.835184654689530\\
35	0.593526580613746\\
};
\addlegendentry{DI, $M/L=2$}

\addplot [color=blue, line width=2.0pt, mark=triangle, mark options={solid, blue}]
  table[row sep=crcr]{%
15	70.7194136624511\\
17.5000000000000	70.6677670883177\\
20	67.9691094555595\\
22.5000000000000	45.0860507581219\\
25	6.04248313572031\\
27.5000000000000	4.76160615291238\\
30	4.47407604956386\\
32.5000000000000	4.07085671866608\\
35	3.82012122901614\\
};
\addlegendentry{DI, $M/L=5$}

\addplot [color=magenta, dashdotted, line width=2.0pt]
  table[row sep=crcr]{%
15	1.35600564529435\\
17.5000000000000	1.01686078122836\\
20	0.762538011540426\\
22.5000000000000	0.571822839250053\\
25	0.428806635917658\\
27.5000000000000	0.321559613197981\\
30	0.241135691892351\\
32.5000000000000	0.180826259013450\\
35	0.135600564529436\\
};
\addlegendentry{CRB, $M/L=2$}

\addplot [color=green, dashed, line width=2.0pt]
  table[row sep=crcr]{%
15	1.27618135247515\\
17.5000000000000	0.957001006279170\\
20	0.717649512934089\\
22.5000000000000	0.538161214079528\\
25	0.403563978125561\\
27.5000000000000	0.302630290291528\\
30	0.226940702258217\\
32.5000000000000	0.170181518485279\\
35	0.127618135247515\\
};
\addlegendentry{CRB, $M/L=5$}

\end{axis}
\end{tikzpicture}%
            \vspace{-0.3cm}
    \caption{RMSE of azimuth estimation of joint and Doppler-ignorant (DI) estimator  for $M/L=2$ and $M/L=5$ with CRB comparison.}
    \label{fig:rmse_az}\vspace{-5mm}
\end{figure}

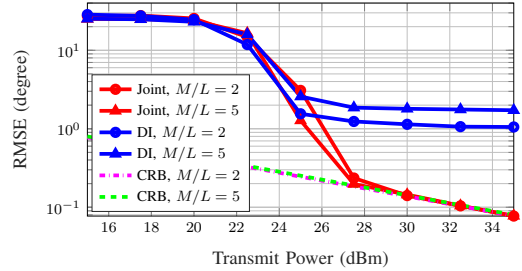
\begin{figure}[t]
    \centering
\definecolor{mycolor1}{rgb}{1.00000,0.00000,1.00000}%
\begin{tikzpicture}[scale=0.8\columnwidth/10cm,font=\footnotesize]
\begin{axis}[%
width=8 cm,
height=4 cm,
at={(1.011in,0.642in)},
scale only axis,
xmin=15,
xmax=35,
xlabel style={font=\color{white!15!black}},
xlabel={Transmit Power (dBm)},
ymode=log,
ymin=0.0771840953331158,
ymax=40,
yminorticks=true,
ylabel style={font=\color{white!15!black}},
ylabel={RMSE (degree)},
axis background/.style={fill=white},
xmajorgrids,
ymajorgrids,
yminorgrids,
legend style={at={(0.01,0.02)}, anchor=south west, legend cell align=left, align=left, draw=white!15!black},
]
\addplot [color=red, line width=2.0pt, mark=o, mark options={solid, red}]
  table[row sep=crcr]{%
15	27.4181337003916\\
17.5	27.7034279552754\\
20	25.4757082462305\\
22.5	14.7604335729241\\
25	3.08230004061025\\
27.5	0.235576907452529\\
30	0.140731900963452\\
32.5	0.104114272334031\\
35	0.0771840953331158\\
};
\addlegendentry{Joint, $M/L=2$}

\addplot [color=red, line width=2.0pt, mark=triangle, mark options={solid, red}]
  table[row sep=crcr]{%
15	25.0560775853996\\
17.5	24.9688733466644\\
20	23.9733074650822\\
22.5	16.5260391551765\\
25	1.27720063571822\\
27.5	0.197767728553806\\
30	0.144806946464979\\
32.5	0.104652331781288\\
35	0.0776752531640305\\
};
\addlegendentry{Joint, $M/L=5$}

\addplot [color=blue, line width=2.0pt, mark=o, mark options={solid, blue}]
  table[row sep=crcr]{%
15	28.4818866529688\\
17.5	27.4602033646127\\
20	24.5985754301128\\
22.5	11.8358784307091\\
25	1.55228482581758\\
27.5	1.23916053012043\\
30	1.13985131937341\\
32.5	1.06405850903315\\
35	1.0536313753432\\
};
\addlegendentry{DI, $M/L=2$}

\addplot [color=blue, line width=2.0pt, mark=triangle, mark options={solid, blue}]
  table[row sep=crcr]{%
15	25.242390477554\\
17.5	24.8359240370245\\
20	23.0608821540607\\
22.5	15.9408049268573\\
25	2.57436413643655\\
27.5	1.85890064556939\\
30	1.79742191047902\\
32.5	1.76027777841653\\
35	1.72166453425999\\
};
\addlegendentry{DI, $M/L=5$}

\addplot [color=mycolor1, dashdotted, line width=2.0pt]
  table[row sep=crcr]{%
15	0.774735696781442\\
17.5	0.580969812779548\\
20	0.435665898400346\\
22.5	0.32670333441404\\
25	0.244992938646693\\
27.5	0.183718786018497\\
30	0.137769653780859\\
32.5	0.103312665592004\\
35	0.0774735696781442\\
};
\addlegendentry{CRB, $M/L=2$}

\addplot [color=green, dashed, line width=2.0pt]
  table[row sep=crcr]{%
15	0.79839198606479\\
17.5	0.598709527127427\\
20	0.448968807465028\\
22.5	0.336679108888925\\
25	0.252473714159015\\
27.5	0.189328576256503\\
30	0.141976402995911\\
32.5	0.106467282468485\\
35	0.0798391986064785\\
};
\addlegendentry{CRB, $M/L=5$}

\end{axis}

\end{tikzpicture}%
            \vspace{-0.3cm}
    \caption{RMSE of elevation estimation of joint and Doppler-ignorant (DI) estimator  for $M/L=2$ and $M/L=5$ with CRB comparison.}
    \label{fig:rmse_el}
    \vspace{-0.4cm}
\end{figure}

\subsection{Results and Discussion}
To evaluate the performance of the proposed joint estimator, we calculate the RMSE based on Monte Carlo trials and compare it with the CRB. The results for range, Doppler, azimuth, and elevation are given in Figs.~\ref{fig:rmse_range}--\ref{fig:rmse_el}. %\mke{Because of the page restrictions, the figures contain RMSE results for an algorithm called \textit{Doppler-ignorant estimator}, which will be addressed later. }

\subsubsection{Analysis of the Joint Estimator}
\begin{figure}[t]
    \centering
    \input{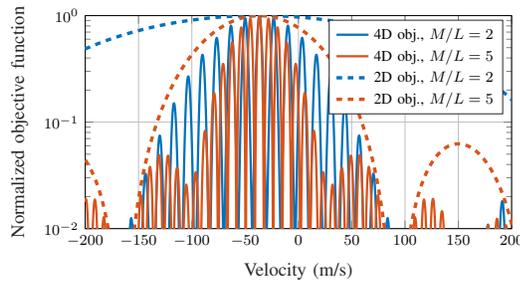}
    \vspace{-0.3cm}
    \caption{{Velocity cut of objective function comparison between $M/L=2$ and $M/L=5$ and for objectives of \eqref{eq:delayDoppEst} and \eqref{eq:glrt_max}. The other parameters are chosen as the exact target values.}}
    \label{fig:obj}
\end{figure}

It is clear %from the RMSE plots that the RMSE 
that the proposed joint estimator shows a waterfall behavior and converges to the CRBs at high SNR, as expected. %At low transmit power values, the RMSEs are upper-limited, due to the finite search ranges. %The reason is that the values of the estimators are constrained. The range is limited to $[0,1250]$ m, the Doppler velocity is limited to $[-321.4,321.4]$ m/s, the azimuth is limited to $[-180,180]$ degrees and the elevation is limited to $[0,90]$ degrees. 
We see that the RMSE of the range behaves similarly for both $M/L=2$ and $M/L=5$. This is because the frequency-domain steering vector remains unaffected by RIS phase profiles. 
However, for Doppler and angle estimation, we observe that the RMSE for $M/L=2$ converges to the CRB later than that for $M/L=5$ in Figs.~\ref{fig:rmse_dopp}--\ref{fig:rmse_el}. This results from the nature of our estimator and can be explained by contrasting the original 4D objective functions \eqref{eq:glrt_max} for the studied $M/L$ values, obtained by the repetitive RIS phase profiles. 
The Doppler cuts of both 4D and 2D objective functions \eqref{eq:glrt_max} and \eqref{eq:delayDoppEst} are shown in Fig.~\ref{fig:obj} for different $M/L$. 
The 4D objective function has several local maxima that are closer and has smaller amplitude when $M/L$ is higher. In other words, the sidelobes are higher and are at a larger distance when $M/L$ is smaller. Hence, the other local maxima can exceed the main lobe in the presence of noise, and the estimates can converge to these points, resulting in some outliers, hence, large RMSEs.

%\mke{We have obtained the RMSE results and the results are reasonable. However, to see if our estimator performs adequately, we need some benchmarks. In the following sections, we provide two benchmarks. The first one is a comparison with Doppler-ignorant estimator. The other one is a benchmark on the effect of $M/L$ on estimation where we observe the performance of the estimator for different $M/L$ values with the same observations. }

\subsubsection{Comparison with Doppler-Ignorant Approach}

%\mke{To see the effects of the Doppler estimation, we compare the joint estimator with a Doppler ignorant estimator. In the Doppler ignorant case, we apply only IFFT on the frequency domain to get the range profile for each repetition interval. To increase the estimation performance, we noncoherently integrate over time domain for each range bin, and pick the largest point as the range estimation. For better estimation, perform gradient search by using the \eqref{eq:delayDoppEst} with Doppler is fixed to zero. Then, using the estimated range and fixed Doppler value, azimuth and elevation is searched over the complete azimuth-elevation domain. At this search, a grid of points with 2-degree step size in azimuth-elevation is created. We choose the maximum point and perform gradient search by using \eqref{eq:angleEst}. Using these estimations and again fixing the Doppler at zero, we perform gradient search with objective \eqref{eq:GLRT_4D} in by initially starting at the estimations for fine-tuning.}

{When applying the DI estimator, we see that range estimation in Fig.~\ref{fig:rmse_range} is slightly degraded, due to biases induced by the Doppler. 
%With the simulation results for both joint and Doppler-ignorant estimator that are shown in Figs. \ref{fig:rmse_range},\ref{fig:rmse_az},\ref{fig:rmse_el}, we can observe how we can perform better estimation jointly than the Doppler ignorant case. It can also be observed that range estimation can converge to a level. However, since there is a bias in the time domain, which is caused by ignoring the Doppler effect, it also increases the RMSE level of range estimation but with the same characterization as the joint estimator. 
On the other hand, the RMSE of angle estimation in Fig.~\ref{fig:rmse_az}--\ref{fig:rmse_el}  shows a different trend and is more severely affected by the Doppler effect. This is due to angle-Doppler coupling, as both parameters are estimated by exploiting phase variations across time. % unlike the range. Additionally, with some other results that we cannot put because of the page restrictions, we saw that the performance of the Doppler ignorant case depends on the Doppler value. If the Doppler is too small, for example, 1m/s, the Doppler-ignorant estimator beats the joint estimator until very high SNR values are reached. Because the difference between the true Doppler and zero is so small that it is lower than the Doppler estimation error of the joint estimator.}
When we compare the impact of $M/L$, the joint and Doppler-ignorant algorithms show  different trends: joint estimation benefits from larger $M/L$, while Doppler-ignorant estimator is better at smaller $M/L$. 
 Normally, it is expected that the performance of angle estimation would be enhanced for smaller $M/L$ since we have more degrees of freedom (i.e., more diversity in RIS phase profiles), as in the case of the DI estimator. However, in Figs.~\ref{fig:rmse_az}--\ref{fig:rmse_el}, we have the opposite situation for the joint estimator since the angle error is affected by the error of Doppler estimation, which is shown in Fig.~\ref{fig:rmse_dopp}. It is observed that for high transmit powers, the Doppler estimation error is higher for $M/L=2$. This is expected since for $M/L=2$, the 4D objective function has higher sidelobes at larger distances as shown in Fig. \ref{fig:obj}. Therefore, the joint estimator leads to larger RMSEs on angle estimation for smaller $M/L$, resulting from the poor Doppler estimation in \eqref{eq:delayDoppEst} (which in turn deteriorates the angle estimation performance in \eqref{eq:angleEst} and affects the initialization point of 4D gradient ascent in \eqref{eq:GLRT_4D}).}

\subsubsection{Detection Performance}
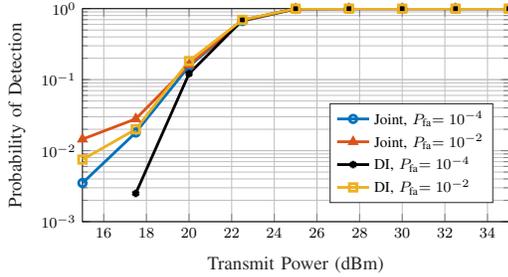
\begin{figure}[t]
    \centering
\definecolor{mycolor1}{rgb}{0.00000,0.44700,0.74100}%
\definecolor{mycolor2}{rgb}{0.85000,0.32500,0.09800}%
\definecolor{mycolor3}{rgb}{0.92900,0.69400,0.12500}%
\begin{tikzpicture}[scale=0.8\columnwidth/10cm,font=\footnotesize]
\begin{axis}[%
width=8 cm,
height=4 cm,
at={(1.011in,0.642in)},
scale only axis,
xmin=15,
xmax=35,
xlabel style={font=\color{white!15!black}},
xlabel={Transmit Power (dBm)},
ymode=log,
ymin=0.001,
ymax=1,
yminorticks=true,
ylabel style={font=\color{white!15!black}},
ylabel={Probability of Detection},
axis background/.style={fill=white},
xmajorgrids,
ymajorgrids,
yminorgrids,
legend style={at={(0.58,0.088)}, anchor=south west, legend cell align=left, align=left, draw=white!15!black}
]
\addplot [color=mycolor1, line width=1.5pt, mark=o, mark options={solid, mycolor1}]
  table[row sep=crcr]{%
15	0.0035\\
17.5	0.018\\
20	0.1545\\
22.5	0.6685\\
25	0.993\\
27.5	0.9995\\
30	1\\
32.5	1\\
35	1\\
};
\addlegendentry{Joint, ${P}_{\text{fa}}{=10}^{{-4}}$}

\addplot [color=mycolor2, line width=1.5pt, mark=triangle, mark options={solid, mycolor2}]
  table[row sep=crcr]{%
15	0.0145\\
17.5	0.028\\
20	0.161\\
22.5	0.672\\
25	0.993\\
27.5	0.9995\\
30	1\\
32.5	1\\
35	1\\
};
\addlegendentry{Joint, ${P}_{\text{fa}}{=10}^{{-2}}$}

\addplot [color=black, line width=1.5pt, mark=asterisk, mark options={solid, black}]
  table[row sep=crcr]{%
15	0\\
17.5000000000000	0.00250000000000000\\
20	0.121250000000000\\
22.5000000000000	0.686250000000000\\
25	0.998750000000000\\
27.5000000000000	1\\
30	1\\
32.5000000000000	1\\
35	1\\
};
\addlegendentry{DI, ${P}_{\text{fa}}{=10}^{{-4}}$}

\addplot [color=mycolor3, line width=1.5pt, mark=square, mark options={solid, mycolor3}]
  table[row sep=crcr]{%
15	0.00750000000000000\\
17.5000000000000	0.0200000000000000\\
20	0.182500000000000\\
22.5000000000000	0.695000000000000\\
25	0.998750000000000\\
27.5000000000000	1\\
30	1\\
32.5000000000000	1\\
35	1\\
};
\addlegendentry{DI, ${P}_{\text{fa}}{=10}^{{-2}}$}

\end{axis}

\end{tikzpicture}%
    \vspace{-0.3cm}
    \caption{Probability of detection versus transmit power for two different values of probability of false alarm.}
    \label{fig:roc_snr}
    \vspace{-0.5cm}
\end{figure}

{Besides the estimation performance, we %obtain ROC curves to 
investigate the detection performance with respect to the transmit power for several false alarm rates, which is shown in Fig.~\ref{fig:roc_snr}. For a specified false alarm rate, we choose a threshold that ensures the same false alarm rate on the noise-only measurements. We observe that the detection performance is around one for transmit powers higher than 25 dBm. Additionally, it is seen that the joint estimator outperforms the Doppler ignorant one. Hence, we can conclude that it is crucial to account for the Doppler effect.}%This is observed that for a suitable detection rate, transmit power should be at least $22.5$ dBm, which can be suitable for commercial applications. }

%%%%%%%%%%%%%%%%%%%%%%%%%%%%%%%%%%%%%%%%%%%%%%%%%%%%%%
\section{Conclusion}

In this work, we addressed the time-domain coupling of Doppler-angle in an \ac{NLoS} RIS-aided monostatic sensing scenario. To solve the coupling, we designed  repetitive RIS phase profiles and accordingly derived a low-complexity target parameter estimator that estimates delay, Doppler and angles based on a GLRT detector. %Then, we derived the GLRT for target detection. 
% We generated a simulation environment for a specified single-target case to analyze its performance. We evaluated the RMSE performance of the proposed GLRT-based low-complexity estimator and 
Simulation results demonstrated the superior performance of the proposed detector/estimator over the state-of-the-art Doppler-ignorant baseline. % constructed the RMSE figures to show that the RMSE converges to the CRBs for sufficiently transmit powers. 
%In addition, the detection performance was investigated to show that with higher transmit powers, we can detect the presence of the target with a reasonable false alarm rate.
Future research will focus on multi-target scenarios.

%%%%%%%%%%%%%%%%%%%%%%%%%%%%%%%%%%%%%%%%%%%%%%%%%%%%%%
%\section*{Acknowledgments}
%This work was supported by Hexa-X-II, part of the European Union’s Horizon Europe research and innovation programme under Grant Agreement No 101095759.

%%%%%%%%%%%%%%%%%%%%%%%%%%%%%%%%%%%%%%%%%%%%%%%%%%%%%%
\balance 
\bibliographystyle{IEEEtran}
\bibliography{IEEEabrv,Sub/main}

\end{document}